\documentclass[12pt,a4paper,reqno]{amsart}

\pdfoutput=1

\usepackage{algorithmic}
\usepackage{url}
\usepackage{fancyhdr}
\usepackage{graphicx}

\newcommand{\pr}[2][]{\ensuremath{\mathrm{Pr}_{#1}\!\left(#2\right)}}

\def \R{\mathbb{R}}

\def \intersect{\cap}

\newcommand{\treespace}{\ensuremath{\mathcal{T}_N}}
\newcommand{\orthant}{\ensuremath{\mathcal{O}}}
\newcommand{\para}{\ensuremath{\parallel}}

\begin{document}

\title[Principal geodesics in treespace]{An algorithm for constructing principal geodesics in phylogenetic treespace}

\author{Tom~M.~W.~Nye}
\email{tom.nye@ncl.ac.uk}
\address{School of Mathematics and Statistics\\
        Newcastle University\\ Newcastle upon Tyne\\ NE1 7RU\\ UK}

\date{May 2014}

\maketitle

\begin{abstract}
Most phylogenetic analyses result in a sample of trees, but summarizing and visualizing these samples can be challenging. 
Consensus trees often provide limited information about a sample, and so methods such as consensus networks, clustering and multidimensional scaling have been developed and applied to tree samples. 
This paper describes a stochastic algorithm for constructing a principal geodesic or line through treespace which is analogous to the first principal component in standard Principal Components Analysis. 
A principal geodesic summarizes the most variable features of a sample of trees, in terms of both tree topology and branch lengths, and it can be visualized as an animation of smoothly changing trees. 
The algorithm performs a stochastic search through parameter space for a geodesic which minimises the sum of squared projected distances of the data points. 
This procedure aims to identify the globally optimal principal geodesic, though convergence to locally optimal geodesics is possible. 
The methodology is illustrated by constructing principal geodesics for experimental and simulated data sets, demonstrating the insight into samples of trees that can be gained and how the method improves on a previously published approach. 
A java package called GeoPhytter for constructing and visualising principal geodesics is freely available from \url{www.ncl.ac.uk/~ntmwn/geophytter}.
\end{abstract}

\maketitle

\vspace{0.3cm}
\begin{center}
\textit{
This is a postprint of an article  published in IEEE Transactions in Computational Biology and Bioinformatics}
\end{center}
\begin{center}
\copyright IEEE/ACM, 2014. 
This is the author's version of the work. It is posted here by permission of ACM for your personal use. Not for redistribution. The definitive version was published in IEEE Transactions in Computational Biology and Bioinformatics, {Vol.\ 11, March-April 2014}, \url{http://doi.ieeecomputersociety.org/10.1109/TCBB.2014.2309599}\\
\end{center}

\section{Introduction}

Samples of phylogenetic trees arise in many different contexts in phylogenetics: examples include Bayesian posterior samples, bootstrap samples and collections of trees from different genetic loci. 
Understanding and summarizing the information present in these samples is challenging due to the difficulty of representing and visualizing regions of the space of possible trees. 
Consensus trees are typically used to summarize samples. 
These represent the features on which trees in the sample tend to agree, and each edge is usually labelled with the proportion of trees in the sample containing that edge. 
Many different consensus methods have been proposed, differing in the particular features considered and the way conflicts between trees are resolved \cite{bry03}. 
For example, the Adams consensus tree \cite{adam72,adam86} is constructed using the relationship between every triplet of leaves represented in a sample of rooted trees, and it does not take branch lengths on the trees into account. 
Alternatively, the average consensus tree \cite{lap97} is defined using the matrix of path length distances between pairs of leaves on each tree, and it therefore depends on branch lengths.  
By definition, consensus trees only reflect limited aspects of the input sample. 
Specifically, they do not indicate alternative tree topologies; they cannot capture correlations between different features; and many do not incorporate information about variability in branch lengths. 
Consensus networks \cite{hol03} were developed to address the first of these issues. 
They represent conflicting phylogenetic signals from different trees in a single network structure. 
Standard tools from multivariate data analysis, such as clustering \cite{sto02,kop07} and multidimensional scaling \cite{hill05} have also been adapted and applied to samples of trees as a means of representing information in the samples more accessibly. 

This paper presents a complementary approach which is analogous to a form of principal components analysis (PCA) adapted to the geometry of the space of phylogenetic trees. 
Given a sample of different phylogenies sharing the same set of taxa, we construct a principal geodesic, or line, in treespace which is a `best fit' to the data in a well-defined sense. 
The principal geodesic can be thought of as a $1$-parameter continuous family of trees which summarizes the sample, as opposed to the point summary provided by a consensus tree. 
In common with consensus network methods, the principal geodesic can represent a collection of alternative tree topologies, depending on the input sample. 
By analogy with the first principal component in standard PCA, the principal geodesic represents the most variable features of the sample, in terms of both topology and branch length. 
As for standard PCA, a `proportion of variance' summary statistic can be computed for each principal geodesic. 
This measures the amount of variability captured by the principal geodesic in relation to the degree of scatter around the geodesic, and therefore indicates how well the principal geodesic summarizes the sample. 

Our approach relies heavily on geometrical properties of the space of phylogenetic trees. 
The space of all possible phylogenetic trees on a fixed set of taxa forms a geometric space which we refer to as \textit{treespace} \cite{bill01}. 
The space is equipped with a metric, usually called the \textit{geodesic metric}, and any pair of points can be joined by a unique \textit{geodesic} i.e.\ a path of minimal length. 
A variety of other metrics exist for measuring differences between phylogenetic trees. 
However, the geodesic metric and the picture of treespace provided by Billera et al \cite{bill01} provide the geometry necessary to perform an analog of PCA: namely, a notion of a straight line and a well-defined projection operation onto any given line.  
Other metrics do not provide these geometrical elements. 

A sample of trees can be regarded as being drawn from some distribution on treespace, and our approach attempts to characterize this distribution by approximating the sample with a principal geodesic. 
An important feature of the approach is that it uses branch length information as well as topological information from the input sample of trees, since information about the shape of the distribution in treespace would be lost by marginalising out the branch lengths. 
For example, if most of the sampled trees had the same topology, then variability in the sample would largely be comprised of variability in the branch lengths for that topology. 
Our approach can characterise the variability equally well in this situation, or conversely for samples which are widely dispersed over different topologies, and it therefore offers a unified approach independent of the nature of the sample. 
Branch lengths and tree topology are intimately related since (i) tree topology can be changed by continuously shrinking edges to length zero and expanding out alternative edges, and (ii) the relative branch lengths in a species tree affect the inferred topology in various ways when constructing a phylogeny from genetic data. 
For example, short internal edges can be harder to infer and so lead to alternative topologies. 
Although biologists are often primarily concerned with the different tree topologies represented within a sample, incorporating branch lengths leads to a more complete picture. 

Although normally applied to data in vector spaces, PCA can be adapted to work in other geometrical spaces. 
Most importantly, PCA has been reformulated in terms of geodesic geometry on Riemannian manifolds \cite{flet04,huck06} including Lie groups and shape spaces \cite{flet03,huck09}.
These ideas have also been extended to certain spaces of trees \cite{wang07}. 
Our algorithm is based on the same type of reformulation, and the first step is to re-express standard PCA in terms of the following schema.  
Suppose $x_1,x_2,\ldots,x_n$ is a set of points in a vector space equipped with an inner product and induced metric $d(\cdot,\cdot)$. 
The zero-th order principal component can be defined as the point $\theta_0$ which minimizes the sum of squared distances $\sum d(x_i,\theta_0)^2$. 
Similarly, the first order principal component $\theta_1$ is the line which minimizes $\sum d(x_i,P_1x_i)^2$ where $P_1$ denotes projection onto $\theta_1$. 
This definition extends to the $k$-th order principal component $\theta_k$ which is the $k$-dimensional subspace which minimizes the sum of squared projected distances $\sum d(x_i,P_kx_i)^2$ where $P_k$ is projection onto $\theta_k$.  
Algebraic solutions can be obtained for these minimization problems for PCA in vector spaces. 
In particular $\theta_0$ is just the sample mean $\bar{x}$, and $\theta_1$ is the line through $\bar{x}$ spanned by the eigenvector of the sample covariance matrix with largest eigenvalue. 
This schema for constructing $\theta_0$ and $\theta_1$ applies exactly as stated above to more general metric spaces equipped with some notion of lines and projection onto lines, such as treespace. 
In terms of treespace geometry, $x_1,\ldots,x_n$ is a sample of trees, $\theta_1$ is a geodesic in treespace and $d(\cdot,\cdot)$ is the geodesic metric. 
However, in the more general setting, algebraic solutions are not available and $\theta_0$ is not generally a subspace of $\theta_1$ \cite{marr10}. 
This can be demonstrated by simple examples in treespace. 
In the context of general metric spaces, the zero-th order principal component $\theta_0$ is more commonly known as the \textit{Fr\'echet mean}, and a stochastic algorithm for computing the Fr\'echet mean in treespace has recently been developed \cite{bac12,mil12} based on previous work in more general metric spaces \cite{sturm03}. 
In this paper we present a stochastic algorithm for constructing $\theta_1$ in treespace, or more precisely, a geodesic which minimises the sum of squared distances between points in the sample and their projections onto the geodesic.  
The algorithm searches through the set of geodesics for a optimal fit to the data, though convergence to local minima is possible. 
We call the algorithm \textit{GeoPhytter}. 

The present author has previously published an algorithm called $\Phi$PCA for constructing the first principal component in treespace \cite{nye11}, and the methods presented here form an updated approach to the same problem, overcoming a number of limitations of the original algorithm. 
Most importantly, $\Phi$PCA only considers geodesics which lie in a restricted class with simple geometrical properties as candidates for $\theta_1$. 
The algorithm fails to construct the optimal geodesic for certain data sets on account of this restriction, and in some cases $\Phi$PCA fails to produce any output at all, such when all the input trees have the same topology. 
Later in the paper we give examples of biological data sets on which $\Phi$PCA does not identify an optimal principal geodesic, and give a more thorough technical explanation of the improvements GeoPhytter represents. 
Other authors \cite{fer13} have more recently also described an algorithm for constructing an approximate principal geodesic in treespace, for which the geodesic is constrained to lie  between points in the data set. 
We apply this algorithm to experimental data sets later in the paper to compare with results obtained using GeoPhytter.   

The remainder of the paper has the following structure. 
After reviewing the geometry of phylogenetic treespace, we then present the stochastic algorithm for constructing principal geodesics. 
Detailed results are given for two data sets: (i) a sample of gene trees for a set of archaea for which gene conversion has affected certain loci and (ii) a simulated bootstrap sample of trees affected by long branch attraction. 
These illustrate potential biological applications of principal geodesic analysis in treespace. 
We also give very brief results for some other data sets to indicate the type of characterization of tree samples that the analysis can provide. 
Finally, we conclude with a discussion of computational issues and possibilities for further research.

\section{Methods}

\subsection{Geometry in treespace}\label{sec:geom}

We need to describe the geometry of phylogenetic treespace in order to specify our algorithm fully. 
More details about the structure and geometry of treespace are given by Billera et al \cite{bill01}, and the account presented here is brief. 
A phylogenetic tree represents the evolutionary relationships between a set of objects, called \emph{taxa}, which label the leaves of the tree.  
Treespace $\treespace$ is the set of all unrooted trees on the set of taxa $S=\{1,2,\ldots,N\}$. 
The trees have positive-valued edge weights or \textit{branch lengths}. 
Sometimes it is convenient to ignore the branch lengths on a tree, in which case we obtain the tree \emph{topology}. 
Topologies are referred to as \emph{resolved} if all the vertices apart from the leaves have degree $3$, but are otherwise called \emph{unresolved}. 
A \textit{split} is a bipartition of $S$, and every edge in an unrooted tree is associated with the split of taxa induced by cutting the edge. 
Splits are often written in the notation $abc\ldots|xyz\ldots$ where $\{a,b,c,\ldots\}\cup\{x,y,z,\ldots\}$ is a disjoint union of $S$.
Although we work throughout with unrooted trees, the space of rooted trees can be obtained by adding in an additional taxon representing the ancestor of all the taxa in $S$, and so the space of rooted trees is essentially isomorphic to $\mathcal{T}_{N+1}$. 
Each tree in $\treespace$ contains up to $N-3$ internal edges and exactly $N$ \textit{pendant} edges, i.e.\ those which contain a leaf. 
Treespace consists of different regions, each corresponding to the set of trees with a particular fully-resolved topology. 
If we consider a single fully-resolved tree topology $\tau$, and ignore the lengths of the pendant edges, then by associating each internal edge with a coordinate axis in $\R^{N-3}$ there is a bijection between the set of trees with topology $\tau$ and the interior of the \textit{positive orthant} $\orthant_\tau=\R^{N-3}_+$ where $\R_+=\{x\in\R:x\geq0\}$. 
The faces of $\orthant_\tau$ correspond to unresolved trees. 
Continuing to ignore pendant edge lengths, it follows that as a set, $\treespace$ corresponds to $ \bigcup \orthant_\tau$, where the union is over all possible fully-resolved tree topologies. 

The orthants corresponding to different topologies overlap along their faces in the following way. 
The codimension-$1$ faces of each orthant correspond to trees in which a single internal edge has contracted to length zero, resulting in a vertex with degree $4$.
Such a vertex can be resolved in three different ways leading to two new topologies $\tau',\tau''$ as well as the original $\tau$. 
It follows that each codimension-$1$ face of $\orthant_\tau$ is identified with corresponding faces in $\orthant_{\tau'}$ and $\orthant_{\tau''}$. 
The topological operation corresponding to shrinking down an edge and replacing it with one of its two alternatives is referred to as \textit{nearest neighbor interchange} (NNI) and adjacent regions in $\treespace$ contain trees related by NNI. 
Higher codimension faces similarly form the intersection of multiple orthants, and the origin of tree space, corresponding to a tree with no internal edges (or \textit{star-tree}), is the unique point in the  intersection of all the orthants. 
Adding back in the pendant edges, $\treespace$ is the product $\R^N_+\times\bigcup \orthant_\tau$ where each point in $\R^N_+$ determines a set of pendant edge-lengths. 

\begin{figure*}
\centering
\includegraphics[scale=0.55]{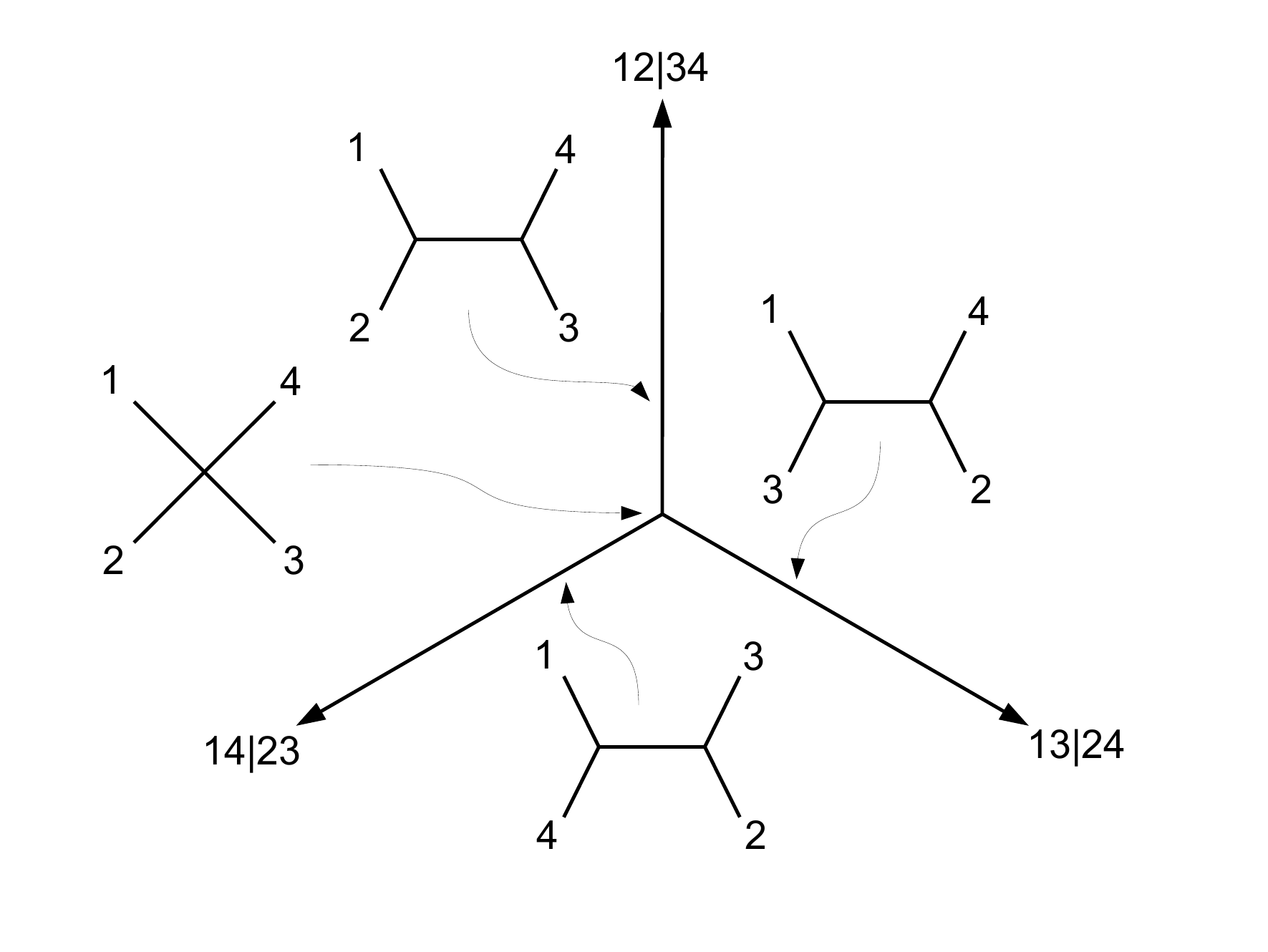}
\caption{
Treespace for $N=4$ taxa consists of three copies of $\R_+$, joined together at a point which corresponds to a star tree. 
Each copy of $\R_+$ is labelled with its corresponding split, and the position point along $\R_+$ determines the length of the edge associated with the split.  
}\label{fig:T4}
\end{figure*}

\begin{figure*}
\centering
\begin{tabular}{cc}
\includegraphics[scale=0.35]{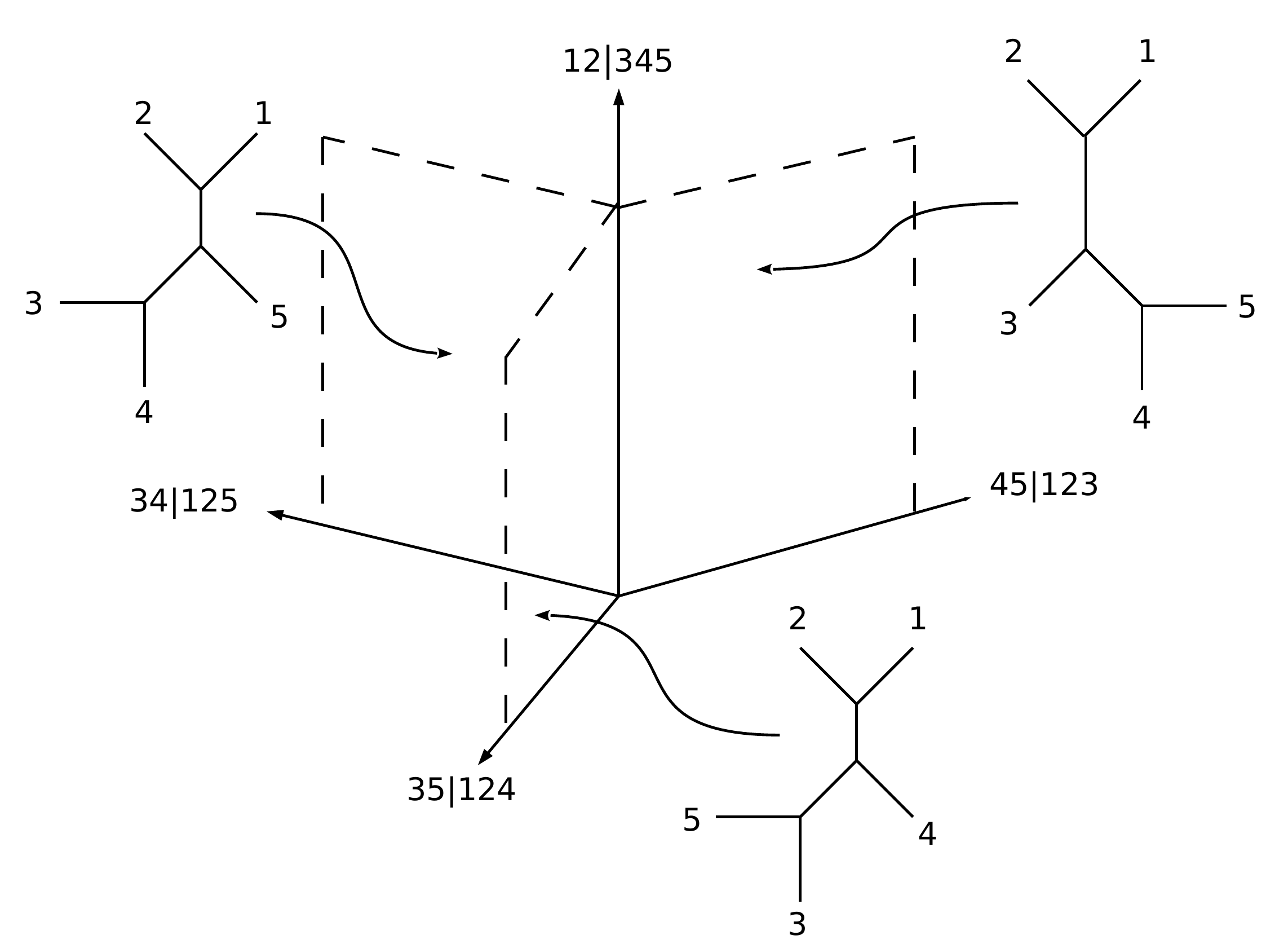} &
\includegraphics[scale=0.45]{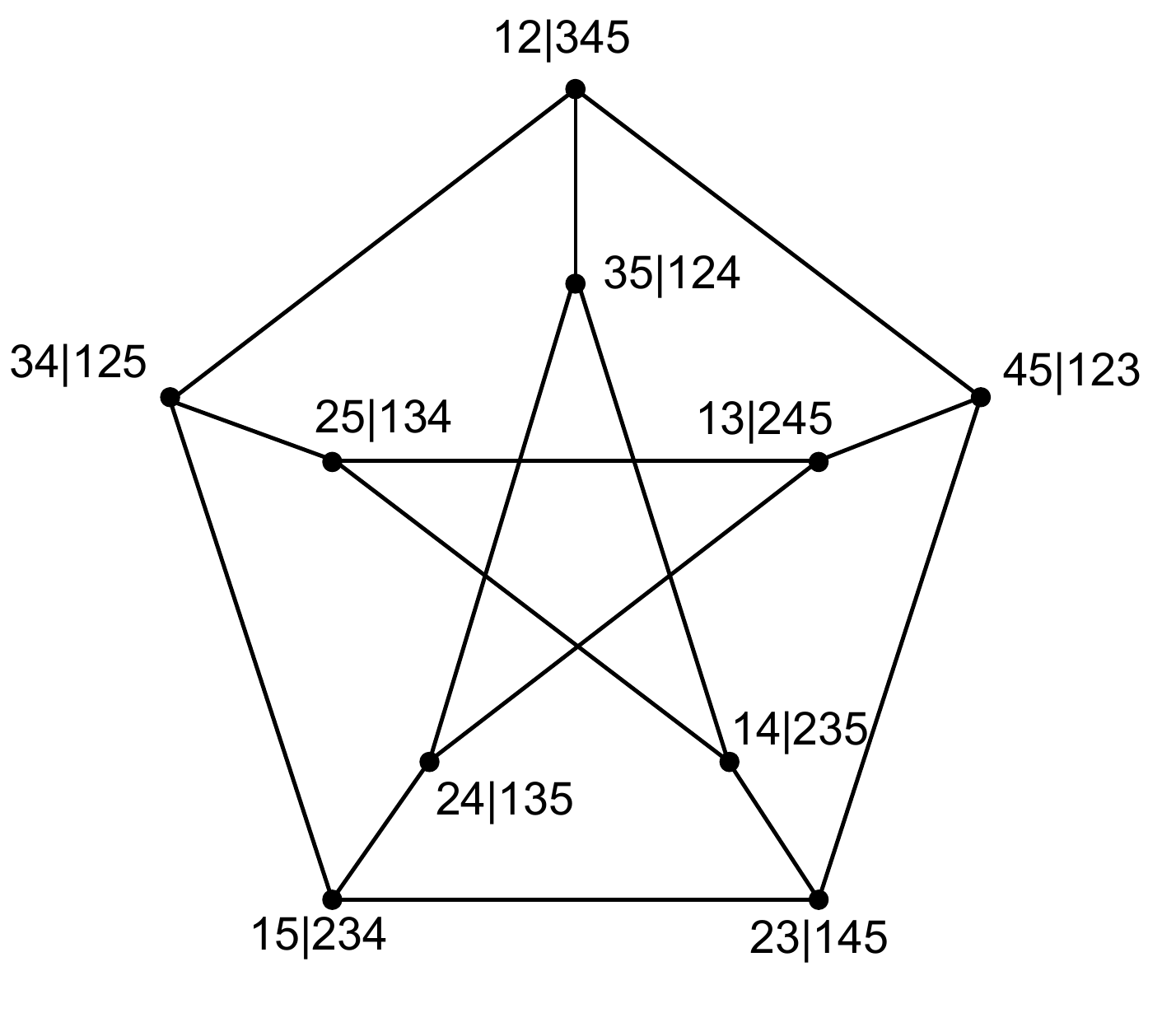}
\end{tabular}
\caption{
Treespace for $N=5$ taxa. 
Left: three orthants in $\mathcal{T}_5$ and their corresponding tree topologies. 
The position within an orthant determines the two internal edge lengths on each tree, and every orthant is glued to two others along each codimension-$1$ face. 
Right: the combinatorial structure of the orthants corresponds to the Petersen graph. 
Each vertex on the graph corresponds to a codimension-$1$ face of an orthant, and is labelled with the split assigned length zero on that face. 
The edges on the graph correspond to the $15$ orthants in $\mathcal{T}_5$. 
The three edges connected to the vertex labelled $12|345$ correspond to the orthants shown on the left.
$\mathcal{T}_5$ is the topological cone of this graph. 
}\label{fig:T5}
\end{figure*}

Insight into the structure of treespace can be gained by considering low-dimensional examples. 
For $N=4$ taxa, there are $3$ possible unrooted tree topologies, so (ignoring pendant edges for simplicity) $\mathcal{T}_4$ consists of three copies of $\R_+$ joined at together at a point which corresponds to a star-tree. 
This is illustrated in Fig.~\ref{fig:T4}. 
For $N=5$ there are $15$ unrooted tree topologies, so $\mathcal{T}_5$ is formed from $15$ orthants each of which is a copy of $\R_+^2$. 
As Fig.~\ref{fig:T5} shows, each orthant is joined to two others along each codimension-$1$ face, corresponding to the two choices for nearest-neighbor interchange on a particular edge of a tree. 
The combinatorial structure of the orthants corresponds to a copy of the Petersen graph, as illustrated by in Fig.~\ref{fig:T5}, and $\mathcal{T}_5$ is the topological cone of this graph. 
For $N=6$ the structure is more complicated again, since unlike the case for $N=4$ and $5$, there are different possible unlabelled tree shapes. 
However, the underlying principles remain the same for all values of $N$: each orthant corresponds to all the trees with a particular topology, and every orthant is joined to two others along each codimension-$1$ face. 

The metric structure on $\treespace$ is obtained as follows. 
Any two trees $x_1,x_2$ with the same topology $\tau$ can be joined by a straight line segment in $\R^N_+\times\orthant_\tau$ and $d(x_1,x_2)$ is defined as the standard Euclidean ($L^2$) length of that segment. 
If $x_1,x_2\in\treespace$ have different topologies then $d(x_1,x_2)$ is defined as the length of the shortest path between $x_1$ and $x_2$ which consists of straight line segments in each orthant, where path length is defined as the sum of individual segment lengths. 
Billera et al\cite{bill01} showed that this determines a well-defined metric $d(\cdot,\cdot)$ on $\treespace$ and that the shortest-length path, or \textit{geodesic}, between any two points is unique.  
A $O(N^4)$ algorithm has been developed for constructing geodesics \cite{owen09}: its input is a pair of trees $x_1,x_2$ and the algorithm outputs a sequence of straight line segments contained in a sequence of orthants linking $x_1$ to $x_2$. 
Any pair of trees $x_1,x_2$ with different topologies $\tau_1,\tau_2$ can be connected by a path consisting of the line segment in $\orthant_{\tau_1}$ joining $x_1$ to the origin and the segment in $\orthant_{\tau_2}$ joining the origin to $x_2$. 
Such paths are called \textit{cone paths} and they sometimes coincide with the geodesic between $x_1$ and $x_2$. 

We use the algorithm of Owen and Provan\cite{owen09} to construct geodesics between points in treespace. 
Given a point $x\in\treespace$ and a geodesic $\gamma$ between two points $\gamma_1,\gamma_2\in\treespace$ there is a unique closest point on $\gamma$ to $x$, which we call the \textit{projection} of $x$, denoted $P_\gamma(x)$. 
Sometimes $x$ is closest to an end-point of $\gamma$ in which case $P_\gamma(x)=\gamma_i$ for $i=1$ or $2$.
The existence of the projection is a result of the particular mathematical structure of $\treespace$ \cite{bill01}, and an efficient algorithm for calculating the projection $P_\gamma(x)$ was presented in \cite{nye11}. 
We use the same algorithm to perform projections here. 
The algorithm involves a path-length parametrization $\gamma(s)$ of $\gamma$, and uses a golden ratio search algorithm to find the value of $s$ which minimises $d(\gamma(s),x)$. 

In contrast to standard PCA, in which the first principal component is an infinite line, in treespace it is advantageous to restrict attention to geodesic segments of finite length. 
To illustrate this, consider a situation in which the projected points $P_\gamma(x_i)$ all lie in a single orthant $\orthant_\tau$ when $\gamma$ has been chosen to minimise the sum of perpendicular squared distances. 
In this case, the extension of $\gamma$ into other orthants is entirely arbitrary, and the data are only informative about the restriction of geodesics to $\orthant_\tau$. 
This situation would arise when the data consist of a tightly clustered collection of trees all having the same topology. 
It follows that more generally, arbitrary samples of trees will not always determine a unique infinite principal geodesic, and so we consider geodesic segments of finite length.  
The suitability of finite geodesic segments as opposed to infinite geodesics for principal geodesic analysis has previously been recognized in \cite{fer13}.

\subsection{Algorithm}\label{sec:alg}

Given a sample of trees $x_1,x_2,\ldots,x_n$, we wish to find a geodesic $\gamma=(\gamma_1,\gamma_2)$ which minimises the objective function:
\begin{multline}\label{equ:obj}
\omega(\gamma;x_1,\ldots,x_n) = \sum_id(x_i,P_\gamma(x_i))^2
\\ +\min_{i} d(\gamma_1,P_\gamma(x_i))^2+\min_{i} d(\gamma_2,P_\gamma(x_i))^2.
\end{multline}
The first term is the sum of squared perpendicular distances, while the last two terms are included so that the global minimum is a geodesic of minimum length. 
The last two terms only contribute to the objective when all the data points project onto the interior of $\gamma$. 
The projection of $x_1,\ldots,x_n$ onto $\gamma$ can be parallelized trivially by partitioning the data set, and our implementation takes advantage of parallel architectures. 
At each step of the algorithm we maintain two points $\gamma_1,\gamma_2\in\treespace$ and the geodesic $\gamma=(\gamma_1,\gamma_2)$. 
We use the notation $\bar{\gamma}_1=\gamma_2$ and $\bar{\gamma}_2=\gamma_1$ so that $(\bar{\gamma}_1,\bar{\gamma}_2)$ represents the same geodesic but with the reverse orientation. 
The algorithm is initialized with $\gamma_1,\gamma_2$ taken to be a random pair of points in the data set.
The algorithm uses a set of stochastic `moves' $f_k$, $k\in K$ which randomly perturb points in treespace. 
Five different moves are available, so $K\subset\{1,2,3,4,5\}$, and the moves are specified below. 
The algorithm operates by repeating the following procedure until convergence is obtained. 

\begin{algorithmic}
\FOR{$k\in K$}
\FOR{$i=1,2$}
\STATE $1)$ Generate a random tree $y=f_k(\gamma_i)$ and construct the geodesic $\eta=(\gamma_i,y)$.
\STATE $2)$ Consider the set of geodesics $\{(\bar{\gamma}_i,z):z\in\eta\}$. 
Use a $1$-dimensional optimization method (golden ratio search) to find $z\in\eta$ which minimises the objective $\omega$ on the restriction to this set. 
\STATE $3)$ Set $\gamma_i=z$.
\STATE $4)$ Either extend or contract $\gamma$ to deal with projection onto the end-points, as detailed below.
\STATE $5)$ The resulting geodesic $\gamma$ is used as the starting point for the next iteration, at step $1$ above. 
\ENDFOR
\ENDFOR
\end{algorithmic}

\begin{figure}[!tpb]
\setlength{\unitlength}{0.3cm}
\centering
\begin{picture}(16,9)

\put(1.0,1.0){\line(1,0){10}}
\put(1.0,1.0){\circle*{0.2}}
\put(11.0,1.0){\circle*{0.2}}

\put(0.3,-0.3){$\bar{\gamma}_i$}
\put(10.0,-0.3){$\gamma_i$}

\put(1.0,1.0){\line(3,1){12}}
\put(11.0,1.0){\line(1,2){3}}

\put(13.0,5.0){\circle*{0.2}}
\put(14,7.0){\circle*{0.2}}

\put(13.7,4.8){$z$}
\put(14.6,7.0){$y$}

\put(15.0,4.6){\vector(1,2){0.75}}
\put(15.0,4.6){\vector(-1,-2){0.75}}

\end{picture}

\caption{Illustration of one step of the algorithm. 
A point $y$ is obtained by applying a stochastic rule $f_k$ to one end $\gamma_i$ of $\gamma$, and the geodesic $\eta=(\gamma_i,y)$ is constructed. 
The objective $\omega$ is evaluated for different geodesics $(\bar{\gamma}_i,z)$ where $z\in\eta$ and $\bar{\gamma}_i$ is the opposite end to $\gamma_i$. 
Each evaluation involves projecting the data points onto the candidate geodesic $(\bar{\gamma}_i,z)$. 
By applying a golden ratio search method, $\gamma_i$ is replaced by the value of $z$ which minimises the objective. 
}\label{fig:alg}
\end{figure}
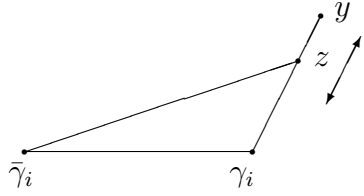

An illustration of the procedure is given in Fig.~\ref{fig:alg}. 
The moves $f_1,f_2,\ldots,f_5$ for randomly perturbing the end-points of $\gamma$ are specified as follows.

\vspace{1ex}
\noindent\textbf{Gaussian random walk:} 
Given an initial fully-resolved tree $y_0=x$ we will define a random sequence of trees $y_1,y_2,\ldots, y_{n_{\text{RW}}}$ for some fixed number of steps $n_{\text{RW}}$ and take $f_1(x) = y_{n_{\text{RW}}}$. 
To define this sequence, we fixing an ordering of the internal edges $e_1,\ldots,e_{N-3}$ in $x$ and let $l(e)\in\R^+$ denote the length of $e$. 
To obtain $y_{j+1}$ from $y_{j}$ we consider each internal edge $e$ in $y_j$ in turn and apply the follow procedure:
\begin{enumerate}
\item Generate a normally distributed random variate $z\sim N(0,\sigma^2)$ and let $l^\ast=l(e)+z$.
\item If $l^\ast\geq 0$ set the length of $e$ to be $l^\ast$. 
\item Conversely if $l^\ast<0$ choose one of the two NNI replacements for $e$ uniformly at random, denote it $e'$, and replace $e$ in $y_j$ by $e'$, setting $l(e') = |l^\ast|$. 
\end{enumerate}
We call this a \textit{Gaussian random walk} on treespace, and it provides a straightforward means of randomly perturbing trees. 
For the results in this paper, a fixed number of steps $n_{\text{RW}}=10$ was used for the Gaussian random walk, and $\sigma^2$ was fixed to give a certain approximate probability that each edge was replaced by at least one NNI during the walk. 
If the mean internal edge length in the data set is denoted $\bar{l}$, then the probability of at least one NNI on a given edge is approximately $\pr{X<0}$ where $X\sim N(\bar{l},n_{\text{RW}}\sigma^2)$. 
We took $\sigma^2 = \bar{l}/0.85n_{\text{RW}}$, to give an approximate probability of $20\%$.

\vspace{1ex}
\noindent\textbf{NNI:} $f_2(x)$ is obtained by sampling an internal edge $e$ in $x$ uniformly at random, and randomly selecting an NNI replacement $e'$ for $e$ from the two possibilities. 
The length of $e'$ in $f_2(x)$ is set to a fixed value (the largest edge length observed in the data set). 

\vspace{1ex}
\noindent\textbf{Random data point:} $f_3(x)$ is taken to be a data point $x_i$ sampled uniformly at random from $x_1,\ldots,x_n$. 

\vspace{1ex}
\noindent\textbf{End-point move:} Suppose that data points $x_{r_1},x_{r_2}, \ldots, x_{r_D}$ project onto the end $\gamma_i$ of the geodesic. 
An approximate Fr\'echet mean is computed for these data points via the algorithm of Miller et al \cite{mil12} and Bacak \cite{bac12}, as follows. 
First $x_{r_1},x_{r_2}, \ldots, x_{r_D}$ are randomly permuted, and then we fix $y_1=x_{r_1}$. 
To obtain $y_{j+1}$ from $y_j$ we construct the geodesic $(y_j,x_{r_{j+1}})$ and let $y_{j+1}$ be the point a proportion $1/(j+1)$ along the geodesic. 
This rule is applied iteratively for $j=2,3,\ldots,D$ to obtain $f_4(\gamma_i)=y_{D}$. 
If no points project onto $\gamma_i$ then this move is not performed. 

\vspace{1ex}
\noindent\textbf{Pendant edge lengths:} $f_5(x)$ is obtained by randomly perturbing the pendant edge lengths in $x$. 
Specifically, for each pendant edge $e$ with length $l(e)$, $e$ is assigned a new length drawn from a gamma distribution with mean $l(e)$ and variance $\nu^2$. 
We took $\nu^2=n_{\text{RW}}\sigma^2$ so that this move had comparable variance to the Gaussian random walk. 

The software allows the user to apply the algorithm with any subset $K\subset\{1,\ldots,5\}$ of moves rather than all five, and also specify the exact order in which the moves are applied. 
Move $f_5$ is only required if pendant edges are included in the analysis. 
By default the software uses moves $1$--$3$ applied in index order, and pendant edges are ignored -- the metric is computed without including them. 
However, a software option allows the user to include pendant edges in the analysis together with move $f_5$.

\subsection{Contracting and extending $\gamma$}

Every time $\gamma$ is updated at step $4$ in the algorithm above, $\gamma$ is either extended or contracted in order to reduce the objective $\omega(\gamma)$ further, in the following way. 
Consider the situation when none of the data points project onto a particular end of $\gamma$. 
Without loss of generality, suppose this end is $\gamma_1$. 
Then $\gamma_1$ can be replaced with the closest projected point $P_\gamma x_i$, thereby making the middle term in~$\eqref{equ:obj}$ vanish but leaving the other terms unchanged.
Thus contraction always reduces the objective whenever no points project onto either end of $\gamma$. 
The contraction operation leads to estimated principal geodesics which are of minimal length, and it ensures that at least one data point projects onto each end of $\gamma$ at each step of the algorithm.
 
Conversely, if any data points project onto an end $\gamma_i$ and that end lies in the interior of an orthant, then $\gamma$ can be extended up to the boundary of the orthant and the projection of these data points re-calculated. 
For example, if $\gamma_1$ lies in the interior of $\R^N_+\times\orthant_{\tau_1}$ then since $\gamma\intersect(\R^N_+\times\orthant_{\tau_1})$ is a line segment it has a unique extension either from $\gamma_1$ up to the boundary of $\R^N_+\times\orthant_{\tau_1}$ or an extension out to infinity. 
Extension across boundaries is non-unique, since there is always a choice about which adjacent orthant to extend into, and so we only make use of the unique extension to the boundary.
Now suppose without loss of generality that a data point $x$ projects onto the end $\gamma_1$ and define $\phi(s)=d(x,\gamma(s))$ where $\gamma(s)$ is a linear parametrization of $\gamma$ for $s\in[0,1]$. 
The extension of $\gamma$ in $\orthant_{\tau_1}$ enables the domain of $\phi$ to be extended to an open neighbourhood $s\in(-\epsilon,\epsilon)$ of zero, and the definition of the metric ensures $\phi$ is continuous on this domain. 
Since $x$ projects onto the point $s=0$, $\phi$ decreases as $s$ tends to zero from above. 
It follows that extension and re-calculation of the projection of $x$ will lead to a reduction in the objective except in the special case that $\phi$ has a minimum at $s=0$.   
Note that the shortest possible extension which improves the objective is taken, in the sense that after extension the last two terms of~$\eqref{equ:obj}$ vanish.    
If many points project onto an end-point, and that end-point lies on the face of an orthant (i.e. the tree is unresolved) it might indicate that a geodesic with a better objective value could be obtained by moving the end-point into a neighbouring orthant, and that consequently the geodesic is sub-optimal.  
However, this seems to occur rarely in practice.

\subsection{Convergence and local minima}

The algorithm is not guaranteed to converge to a global minimum. 
In practice the optimization procedure is halted if the fractional change in objective is below a fixed threshold for a large number of iterations dependent on the number of trees in the data set and the number of taxa. 
To test optimality of the resulting geodesic, the algorithm can be run several times, with a different starting value for $\gamma$ each time. 
If multiple different runs converge to very similar geodesics, that suggests a global optimium has been found, though it cannot be guaranteed. 
A unique global minimum for the objective may not exist: that is the case for standard PCA when several eigenvectors of the data covariance matrix share the same maximimal eigenvalue. 
Since the analysis in treespace essentially reduces to standard PCA when all the data points are highly concentrated in a single orthant, treespace also lacks unique optima in general. 

\subsection{Rationale}

The high dimensionality and complex combinatorial structure of treespace motivated the decision to develop a stochastic algorithm to optimize the objective function in equation~$\eqref{equ:obj}$. 
A simple Localised Random Search algorithm \cite{spal03} would have the following form:
\begin{enumerate}
\item Start with a random geodesic $\gamma$.
\item Randomly perturb $\gamma$, for example by performing a Gaussian random walk on one of its ends, to obtain a new geodesic $\gamma^*$.
\item If $\omega(\gamma^*)<\omega(\gamma)$ set $\gamma=\gamma^*$.
\item Repeat from step 2, unless some convergence criterion has been satisfied. 
\end{enumerate}
This algorithm is simple to implement, but its performance depends on the exact nature of stochastic innovation at step 2. 
In general, if an innovation with small variance is used, then the proposed improvement $\gamma^*$ tends to be close to $\gamma$, and the algorithm slowly moves downhill to settle at a local minimum. 
Conversely, by using an innovation with large variance, larger steps are possible enabling the algorithm to traverse between local minima more easily, but at step 3 the proposed improvement $\gamma^*$ is often rejected. 
We tested a version of Localised Random Search for which the innovation consisted of a Gaussian random walk on a randomly chosen end of the geodesic $\gamma$. 
This algorithm performed very badly for a range of different variances in the random walk step, in particular since most proposals $\gamma^*$ failed to improve the objective.
 
GeoPhytter is based on Localised Random Search, but it combines stochastic innovations with deterministic optimization over slices in treespace. 
The combination of the stochastic moves followed by deterministic optimization over slices gave much better performance in terms of computational time and convergence properties. 
In particular, it employs both `local' innovations (like the NNI move) and `global' innovations (namely the random data point move) which enable the algorithm to explore the local neighbourhood of $\gamma$ while still having the potential to escape from local minima. 

The specific moves $f_1,\ldots,f_5$ were designed under the following rationale. 
Moves $f_1$ and $f_2$ represent `local' innovations (although the variance of the random walk can be adjusted to make it perform bigger steps), while $f_3$ enables larger `global' moves. 
Additionally, $f_1,f_2$ enable $\gamma$ to move out of the convex hull of the data set; $f_3$ cannot achieve this on its own, but it is easy to construct examples for which the principal geodesic does not lie in the convex hull of the data. 
Move $f_4$ tends to shift the ends of the segment $\gamma$ outwards in the direction of the set of points projecting onto each end. 
The random perturbation of pendant edges, $f_5$, explores nearby geodesics for which the pendant edge lengths differ from $\gamma$. 
It provides a counterpart to $f_1$ and $f_2$, which do not affect the pendant edges.

The order in which the different moves were applied had little effect on the convergence properties of the algorithm. 
However, the convergence properties did depend on the selection of moves employed. 
Runs of the algorithm which used all the moves tended to converge faster than those using a limited set. 
Move $f_3$, selection of a random data point, is crucial: the algorithm performs relatively well using this move alone, but without it the algorithm converges more slowly and multiple runs often fail to converge to the same geodesic. 

While pendant edges can optionally be included in any analysis, their inclusion tends to worsen convergence. 
In particular, when pendant edges are included the algorithm  converges to local minima more frequently (indicated by multiple runs converging to different geodesics). 
A greater number of replicate runs is therefore required. 
Furthermore, for some data sets we analysed, the variability in pendant edge lengths swamped the signal from internal edges. 
Since biological interest principally lies with internal edges, we recommend that the software is used primarily for analyses ignoring pendant edges.
 
\subsection{Proportion of variance statistic}\label{sec:rsqu}

In standard PCA the sum of squared distances from the sample mean decomposes as
\begin{equation}\label{equ:ssq}
\sum d(x_i,\bar{x})^2=\sum d(x_i,P_1x_i)^2 + \sum d(P_1x_i,\bar{x})^2
\end{equation}
where $P_1$ denotes projection onto the first principal component $\theta_1$. 
The quantity $r^2=\sum d(P_1x_i,\bar{x})^2 / \sum d(x_i,\bar{x})^2$ is usually reported and is interpreted as the proportion of variance in the sample explained by $\theta_1$. 
The Pythagorean theorem does not hold in treespace so a decomposition like~$\eqref{equ:ssq}$ does not apply. 
Nonetheless, an analog of $r^2$ can be defined as follows. 
Let
\begin{equation*}
d_\perp^2 = \sum d(x_i,P_\gamma(x_i))^2,\quad\text{and}\quad d_\para^2 = \sum |s_i-\bar{s}|^2
\end{equation*}
where $s_i=d(\gamma_1,P_\gamma(x_i))$ for $i=1,\ldots,n$ and $\bar{s}$ is  the mean of the $s_i$. 
Specifically $s_i$ is the position along the geodesic $\gamma$ of the projection of $x_i$ obtained by the projection algorithm described in section~\ref{sec:geom}. 
Equivalently, $d_\para^2$ can be defined as $\sum d(y_i,\bar{y})^2$ where $y_i=P_\gamma(x_i)$ and $\bar{y}=\gamma(\bar{s})$ is the Fr\'echet mean of $y_1,\ldots,y_n$. 
The quantity $r^2_\gamma = d_\para^2/(d_\para^2+d_\perp^2)$ is then the analog of $r^2$ in the standard analysis, and it is readily computed for any geodesic $\gamma$.
For standard PCA in $\R^k$, $r^2$ is bounded below by $1/k$, with equality only in the case of isotropic data. 
This inequality does not hold in treespace, but it forms a useful baseline with which to interpret the statistics $r^2_\gamma$, by taking $k=N-3$ (the number of internal edges).  
Note that a geodesic which minimizes $d^2_\perp$ does not necessarily maximize $r^2_\gamma$. 
This is essentially a result of the failure of the Pythagorean theorem in treespace, and examples based on trees containing $4$ taxa can readily be constructed to demonstrate this.

\subsection{Comparison with existing algorithms}

The previously published algorithm, $\Phi$PCA, was based on a competely different method for constructing principal geodesics, by building the principal geodesic up one split at a time. 
More importantly, $\Phi$PCA made certain assumptions about the principal geodesic which we do not make here, as follows:
\begin{enumerate}
\item In $\Phi$PCA, the principal geodesic was forced to lie in a restricted set of geodesics, called \textit{simple} geodesics. 
These have the property that as the geodesic is traversed, at most one edge is shrunk down to have zero length at a time. 
This rules out many possibilities, such as all geodesics which are cone paths, and the restriction was made for computational convenience rather than biological reasons. 
The algorithm presented here makes no such restrictions.
\item $\Phi$PCA sought to construct an infinite principal geodesic; we explained above why the data are only generally informative about a finite principal geodesic segment. 
On account of this, $\Phi$PCA fails to find the principal geodesic in certain cases, for example when the data do not represent every topology that arises along some infinite extension of a finite principal geodesic. 
\item The principal geodesic in $\Phi$PCA was forced to pass through a consensus tree under the assumption that this tree lay close to the Fr\'echet mean. 
The present algorithm does not restrict $\gamma$ to pass through any mean or consensus tree. 
As explained in the introduction, the geodesic which minimises the sum of squared perpendicular distances does not necessarily contain the Fr\'echet mean. 
\end{enumerate} 
The first of these points is probably the most important. 
Given these differences between the algorithms, it is easy to construct simulated data sets for which $\Phi$PCA fails but GeoPhytter successfully identifies the principal geodesic (for example, constructing a data set by simulating trees distributed along a cone-path geodesic and then perturbing each tree slightly via a Gaussian random walk).
In addition, there are biological reasons for believing that for some data sets the principal geodesic will not lie in the class of simple geodesics. 
For example, suppose that trees in some sample of gene trees have one of two topologies which are related by a sub-tree prune and regraft operation (SPR). 
This could arise if a subset of genes were horizontally transferred at some stage on the evolutionary tree. 
Depending on the particular topologies, geodesics between trees with topologies related by SPR are often cone paths, and so it is reasonable to anticipate that for some data sets, the principal geodesic would be a cone path and therefore lie outside the class of simple geodesics. 
The chaperonin data set analysed in the next section supports this intuition: the associated principal geodesics are not contained in the simple geodesic class. 

A stochastic algorithm for constructing an approximate principal geodesic was also presented in \cite{fer13}. 
The geodesic is constrained to lie between a pair of data points $x_i,x_j$. 
Pairs of data points are randomly sampled for a large number of iterations, and the pair which minimizes the sum of squared projected distances is taken as an approximate principal geodesic. 
We compare the performance of this algorithm with GeoPhytter in the next section.

\section{Results}

\subsection{Chaperonin data set}

\begin{figure*}[!tpb]
\centering
\includegraphics[scale=0.65]{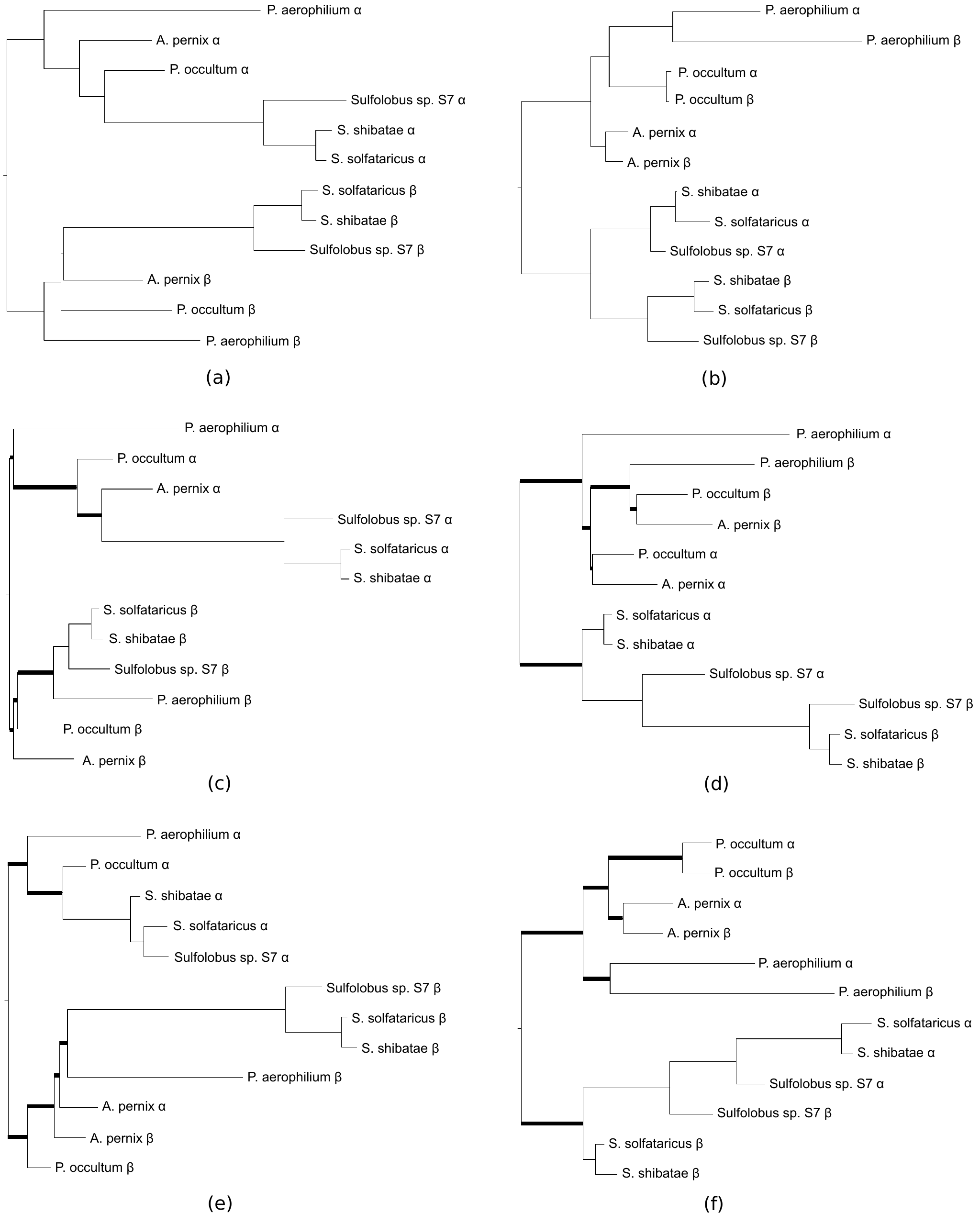}

\caption{
\scriptsize
Phylogenies for the chaperonin data set. 
(a) The duplication topology, and (b) the conversion topology. 
The root splits of these two trees are referred to as the duplication and \textit{Sulfolobus} splits in the main text. 
Leaves are labelled $\alpha$ or $\beta$ according to the copy of the paralog. 
(c) and (d) show the phylogenies at the two ends of the principal geodesic for the sample of trees with window length $200$. 
Edges have been thickened to illustrate the transition between the trees between the phylogenies along the principal geodesic: thickened edges in (c) are simultaneously shrunk to zero length to be replaced by the thickened edges in (d). 
This transition is followed by a change in topology within the \textit{Sulfolobus} clade.
(e) and (f) show the phylogenies at the two ends of the principal geodesic for the sample of trees from the apical domain with window length $100$. 
Again, the thickened edges are simultaneously shrunk to zero length in (e) to be replaced by those in (f). 
There are separate changes in the topology of the \textit{Sulfolobus} clade.
}\label{fig:chap_topos}
\end{figure*}

We applied our method to a sample of trees constructed using an alignment of chaperonin gene sequences from archaea taken from a previous study. 
The original study \cite{arch02} concerned duplication and gene conversion of chaperonin genes, and we analysed the same data set, kindly provided by the authors. 
An ancient gene duplication produced two copies of the chaperonin gene, $\alpha$ and $\beta$, in the archaea included in the study. 
The alignment used to construct phylogenies contained the $\alpha$ and $\beta$ copies of the gene from $6$ different archaea, giving a total of $12$ sequences containing $1557$ DNA sites.
The $6$ archaea were \textit{Pyrodictium occultum}, \textit{Aeropyrum pernix} and \textit{Pyrobaculum aerophilum}, together with $3$ closely related \textit{Sulfolobus} species.
The original analysis suggested two distinct topologies associated with different regions of the alignment. 
The first, which we will refer to as the \textit{duplication topology}, separated the $\alpha$ from the $\beta$ sequences into two highly supported clades, with very similar topologies within the two clades. 
The maximum likelihood tree inferred from the full alignment had this topology. 
In contrast, the second topology, which we will refer to as the \textit{conversion topology}, had separate clades for three organisms (\textit{P.\ occultum}, \textit{A.\ pernix} and \textit{P.\ aerophilum}), each clade containing the $\alpha$ and $\beta$ sequences for the organism. 
This topology was associated with a subset of approximately 300 sites from three separate contiguous loci within a certain domain of the chaperonin protein. 
The topologies are shown in Fig.~\ref{fig:chap_topos}. 
Archibald and Roger \cite{arch02} suggested that multiple independent gene conversion events had given rise to different regions in the gene supporting the two different topologies. 
We call the split separating the $\alpha$ sequences from the $\beta$ sequences the \textit{duplication} split, and conversely, call the split separating the \textit{Sulfolobus} sequences from the others the \textit{Sulfolobus} split.

We constructed phylogenetic trees in a similar way to the original paper, using exactly the same substitution model (general time-reversible plus gamma rate heterogeneity plus invariant sites). 
Maximum likelihood trees were inferred for a series of windows of the alignment each 200 nucleotides long, obtained by progressively sliding the window $10$ nucleotides at a time. 
This gave a sample of $136$ phylogenetic trees. 
The majority consensus of the sample resolved the \textit{Sulfolobus} $\alpha$ and $\beta$ sequences into two separate clades (both of which were present in the duplication and conversion topologies defined above), and contained the \textit{Sulfolobus} split, but was otherwise unresolved. 
A principal geodesic was constructed using the algorithm presented in Sec.~\ref{sec:alg}. 
Pendant edges were ignored in the analysis. 
Ten random starting points for the algorithm converged to the same principal geodesic segment, suggesting that the global optimum geodesic might have been found. 
Exactly one tree in the sample projected onto each end of the segment. 
Fig.~\ref{fig:chap_topos} shows the two trees forming the extremities of the principal geodesic segment.  
The sums of squared distances associated with principal geodesic were $d_\perp^2=27.5$ and $d_\para^2=20.5$ so that $r^2_\gamma=43\%$, indicating that the principal geodesic represents a relatively large proportion of the variability within the sample. 

The principal geodesic is associated with a transition between a topology containing the duplication split and a topology containing the \textit{Sulfolobus} split. 
It essentially represents variability in the relationship between the \textit{Sulfolobus} $\alpha$ and $\beta$ clades. 
The two ends of the principal geodesic do not exactly correspond to the duplication and conversion topologies in the original study, although there is some similarity. 
In particular, the two sequences corresponding to each of the species \textit{P.\ occultum}, \textit{A.\ pernix} and \textit{P.\ aerophilum} are not grouped together in separate clades, as they are in the conversion topology. 
This is not entirely surprising, as the conversion topology was associated with a small proportion of sites in the original alignment, and variability in the phylogenies from the other sites will have played a proportionately higher role in construction of the principal geodesic. 
In addition, the conversion topology was associated with three separate contiguous regions each containing approximately $100$ nucleotides, and so with a sliding window of length $200$ the signal from these regions may have been obscured. 
A principal geodesic was therefore constructed for a sample of trees generated using a sliding window of length $100$, with the windows restricted to cover a particular region, called the apical domain, identified by Archibald and Roger \cite{arch02} as containing the loci affected by gene conversion. 
We call this the \textit{apical} sample. 
This sample contained 50 trees, and the principal geodesic had $d_\perp^2=8.6$, $r^2_\gamma=26\%$. 
Ten runs of the algorithm converged to the same geodesic. 
Fig.~\ref{fig:chap_topos} shows the trees corresponding to the ends of the principal geodesic. 
These have topologies very similar to the duplication and conversion topologies.
The principal geodesic therefore identifies the main source of variability within the sample as coming from the support for these different topologies within the apical domain. 

\begin{figure}[!tpb]
\centering
\includegraphics[scale=0.45]{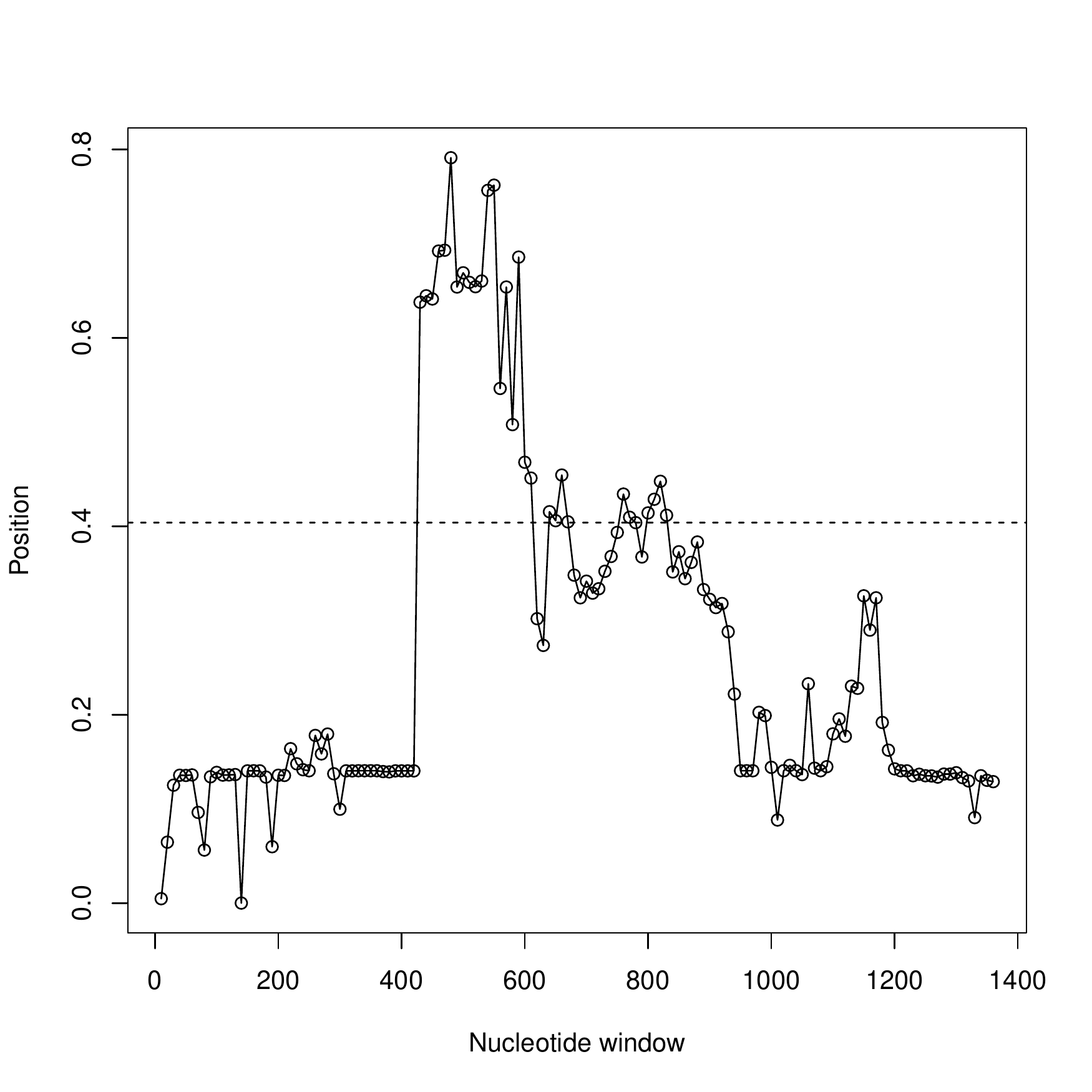} 

\caption{
Projection of the full sample of trees for the chaperonin analysis onto the principal geodesic constructed from the apical sample . 
The $x$-coordinate is the starting point of each window in the alignment, and the $y$-axis is the position of the projected tree along the principal geodesic. 
Small $y$ values correspond to trees near (e) in Fig.~\ref{fig:chap_topos} while large values correspond to trees near (f).
The apical region is approximately between $x=550$ and $x=1050$.
The dashed line corresponds to the tree on the geodesic in which the thickened lines in Fig.~\ref{fig:chap_topos}(e) have been shrunk to zero. 
}\label{fig:proj}
\end{figure}

We computed the projection of the full sample of trees from the sliding window with length $200$ onto the principal geodesic computed for the apical sample. 
The results are shown in Fig.~\ref{fig:proj}. 
The relative position of the projection of a tree along the principal geodesic reflects its similarity to the two alternatives in Fig.~\ref{fig:chap_topos} (e) and (f). 
The peak in the middle of the plot corresponds to a set of trees with the conversion topology. 
The graph acts as a $1$-dimensional summary of the sample: the $y$-coordinate of each tree reflects particular features associated with the tree (for example trees above the dashed line are likely to contain the \textit{Sulfolobus} split, and trees below, the duplication split).
The original analysis of the chaperonin data set\cite{arch02} includes a similar plot, based on a likelihood ratio, for testing relative support for the two conflicting topologies. 
The likelihood ratio approach for testing support is probably advantageous for assessing which regions of an alignment are associated with the two different topologies. 
However, it relies on prior knowledge of those topologies. 
In contrast, score plots like Fig.~\ref{fig:proj} can be prepared for principal geodesics directly from a sample of trees without such knowledge. 

The score plot in Fig.~\ref{fig:proj} contains an apparent cluster of projected points at distance $0.16$ along the geodesic which shows as a horizontal line on the plot. 
This point on the geodesic corresponds to a tree with the same topology as Fig.~\ref{fig:chap_topos}(e) but for which the three \textit{Sulfolobus} $\alpha$ sequences are unresolved, and so it is associated with a change in the topology with the the \textit{Sulfolobus} $\alpha$ clade. 
In order to investigate this effect, we prepared plots of the length of the three possible splits which resolve the \textit{Sulfolobus} $\alpha$ clade versus window position, assigning length zero if the split in question was not present in a tree. 
These plots did not explain the pattern of scores in Fig.~\ref{fig:proj}: specifically, no particular change in split lengths was observed at nucleotide positons $400$ and $950$.  
The full collection of splits and associated lengths for each tree must be taken into account to determine its projection, and it seems as though other features of the principal geodesic beyond these three particular splits determine where trees project relative to the change in the \textit{Sulfolobus} $\alpha$ clade topology on the geodesic. 
In contrast to Euclidean space, the volume of the region which projects to a particular point on a geodesic segment varies from point to point in treespace, and this might be the cause of the apparent clustering of projected points.  
This has implications for principal geodesic analysis and dimensional reduction in treespace more generally, since it is desirable to obtain reductions in which distinct sampled trees are well discriminated. 

Principal geodesics for the full sample and apical sample were also constructed using $\Phi$PCA and the algorithm in \cite{fer13}. 
The algorithm presented in \cite{fer13} was run for $20,000$ iterations. 
This value was chosen since it gave a similar run-time to GeoPhytter.
The results are summarized in Table~\ref{tab:results}. 
$\Phi$PCA produced geodesics with a poor fit to the data. 
The geodesics constructed by GeoPhytter fit the data much better and do not lie in the class of `simple' geodesics required by $\Phi$PCA. 
For example, all the trees on the geodesic constructed by $\Phi$PCA for the apical sample contained the \textit{Sulfolobus} split, so the geodesic failed to represent the variability caused by gene conversion. 
As the table shows, the method of Feragen \textit{et al} performs better than $\Phi$PCA but GeoPhytter is substantially better than both methods in terms of smaller $d^2_\perp$ values and larger $r^2_\gamma$ proportions.

\begin{table}
\begin{center}
\begin{tabular}{c|llll}
& \multicolumn{2}{c}{Full sample}  & \multicolumn{2}{c}{Apical sample}\\
\hline
GeoPhytter & $d_\perp^2=27.5$, &$r_\gamma^2=43\%$ & $d_\perp^2=8.6$, &$r_\gamma^2=26\%$ \\
$\Phi$PCA & $d_\perp^2=39.7$, &$r_\gamma^2=26\%$ & $d_\perp^2=12.8$, &$r_\gamma^2=12\%$ \\
Feragen \textit{et al} & $d_\perp^2=33.5$, &$r_\gamma^2=32\%$ & $d_\perp^2=10.4$, &$r_\gamma^2=15\%$ 
\end{tabular}
\end{center}
\caption{
Comparison of summary statistics for geodesics constructed for the full sample of trees and apical sample using the method presented in this paper (GeoPhytter) and extisting methods. 
}\label{tab:results}
\end{table}

\subsection{Parametric bootstrap data set}

In order to assess the performance of the algorithm on larger trees, we re-analysed a parametric bootstrap sample of trees considered in \cite{nye11}. 
This sample was simulated from an underlying tree containing $41$ taxa representing major eukaryote groups with an outgroup of archaea. 
The tree contained two long branches, corresponding to \textit{microsporidia} and \textit{guillardia}, in addition to the long branch leading to the archaeal outgroup.  
This simulated sample was analysed to explore the possibility of using principal geodesics to capture long branch attraction (LBA) effects. 
The data were transformed prior to construction of principal geodesics by scaling edges: 
edge lengths were scaled so that every split had unit mean length, with the mean for each split taken across trees containing that split in the data set. 
This transformation was performed so that variability was assessed as being relative to edge length, by scaling up the variability in short edges. 
Analysing transformed data in this way can be thought of as being analogous to using the sample correlation matrix instead of the covariance matrix in standard PCA. 
More details are given in \cite{nye11}. 
The analysis was performed ignoring pendant edge lengths in the sample. 

\begin{figure}[!tpb]
\centering
\includegraphics[scale=0.8]{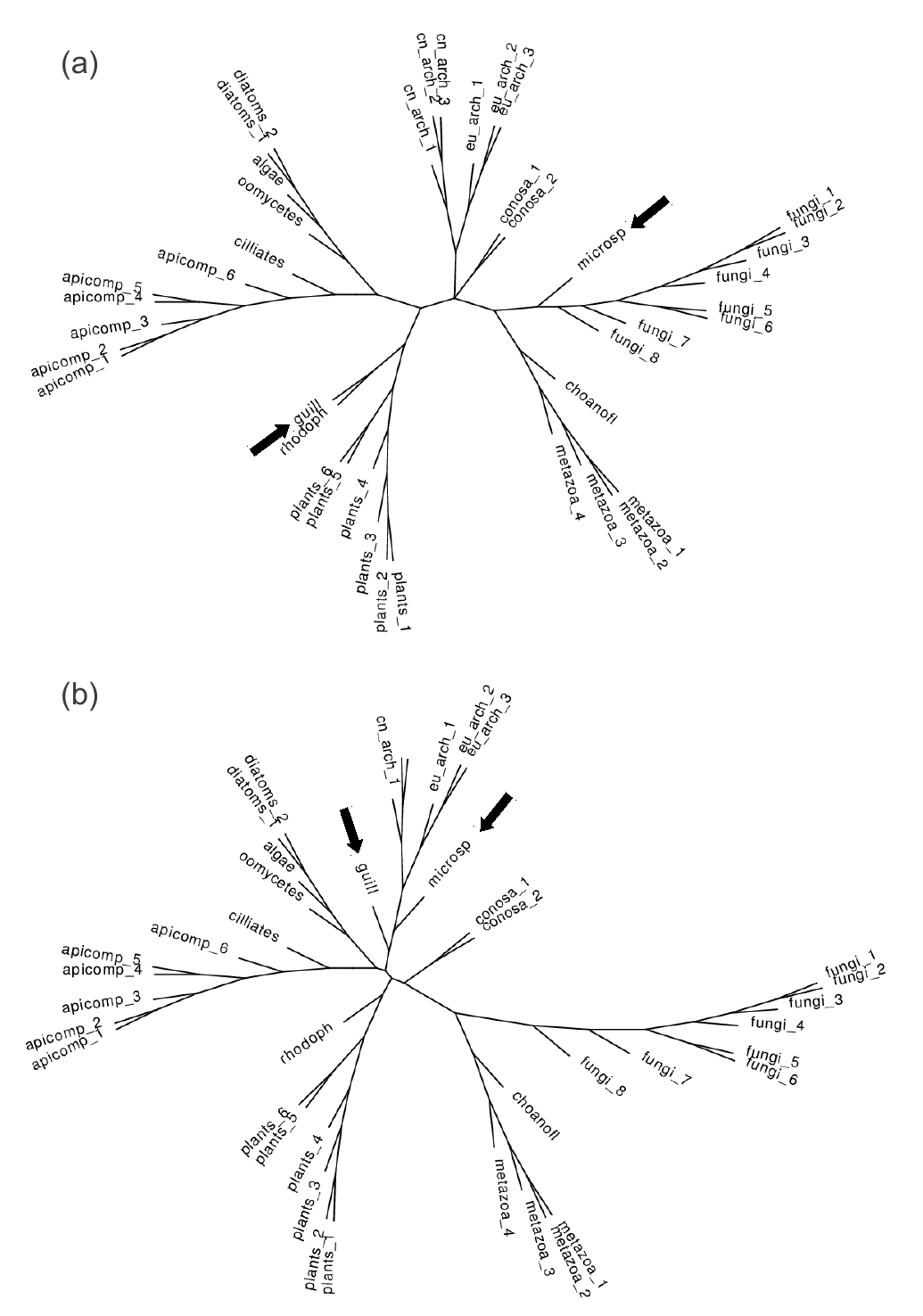}

\caption{
Trees forming the ends of the principal geodesic for the simulated bootstrap sample of trees. 
The arrows mark the taxa \textit{guillardia} and \textit{microsporidia} which have long pendant edges in the original phylogenies. 
The trees have been normalized so that each edge in the data set has unit mean length. 
\textit{Guillardia} and \textit{microsporidia} move around the tree as the principal geodesic is traversed, from their original positions in (a) to be grouped alongside the archaea (labelled `cn arch' and `eu arch') in (b). 
}\label{fig:LBA}
\end{figure}

Fig.~\ref{fig:LBA} shows the end-points of the principal geodesic constructed using the algorithm in Sec.~\ref{sec:alg}. 
As the principal geodesic is traversed, the taxa with long branches, \textit{microsporidia} and \textit{guillardia}, both `float' around the tree from their initial positions near fungi and plants respectively, to be grouped next to the archaea. 
The principal geodesic had $d_\perp^2=1290$ and $r^2_\gamma=13\%$ in comparison to the geodesic constructed with $\Phi$PCA which had $d_\perp^2=1507$ and $r^2_\gamma=10\%$. 
Both \textit{microsporidia} and \textit{guillardia} move around the tree, in contrast to the results from $\Phi$PCA for which only one of the taxa moved in this way.  
The principal geodesic does not lie in the class of simple geodesics considered by $\Phi$PCA, and so $\Phi$PCA is not able to capture the same result. 
When viewed as an animation of trees, the principal geodesic gives an immediate visual representation of the LBA effect present in the data set. 

The value $r^2_\gamma=13\%$ appears smaller than the values typically obtained in a standard principal components analysis. 
This is in part due to the high-dimensional nature of treespace, and the following analogy can be used to interpret $r^2_\gamma$. 
Principal geodesic analysis in treespace with $N$ taxa can be compared with standard PCA in $\R^{N-3}$ when pendant edges are ignored. 
Consider the analysis of multivariate normal data for which the variance is $\sigma^2$ along the principal axis, and $\tau^2$ in all other orthogonal directions. 
It follows that the ratio $\sigma^2 : \tau^2$ is $(N-4)r^2/(1-r^2)$. 
For $r^2=13\%$ and $N=41$ as above this gives a value of $5.5$. 
Under this Euclidean analogy, and assuming the other effects in the data consist of isotropic noise, the principal geodesic for the bootstrap sample has an associated variance which is a factor $5.5$ times the noise variance. 
Of course, the analogy with Euclidean PCA is approximate, but nonetheless this calculation suggests that the LBA effect is significantly greater than random noise.

\subsection{Other data sets}

We constructed principal geodesics for several other data sets in order to assess the performance of the algorithm, and give brief details here to indicate the type of results that can be obtained via principal geodesic analysis. 

\noindent\textbf{Overall scale:} Sometimes gene trees vary in the overall scale of the phylogenies. 
We re-analysed the metazoan data set considered in \cite{nye11}, and obtained a principal geodesic with $r_\gamma^2=57\%$. 
Although this principal geodesic represented some changes in topology, the main feature was the difference in total length of the trees at either end of the geodesic. 

\noindent\textbf{Single NNI:} We analysed a well-known data set of $106$ gene trees for $8$ species of yeast \cite{rok03}, obtained by maximum likelihood inference. 
The principal geodesic passed through the majority consensus topology of the $106$ gene trees and involved a single nearest-neighbor interchange between the subtrees ((\textit{C.albicans}, \textit{S.Kluyveri}), \textit{S.castellii})  and (\textit{C.albicans}, (\textit{S.Kluyveri}, \textit{S.castellii})). 
The principal geodesic had $r^2_\gamma=63\%$, suggesting that the majority of variability in the data set was captured by this single NNI. 

\noindent\textbf{Isotropy:} Some data sets are not well represented by a geodesic segment in tree space. 
In a recent study of turtle evolution \cite{spin13}, phylogenies were inferred for various species of turtle, but the individual representing each species was selected at random from a fixed pool of individuals. 
Phylogenies were constructed for 100 different sets of individuals sampled in this way. 
The principal geodesic constructed for this data set had $r^2_\gamma$ close to the isotropic baseline discussed in Sec.~\ref{sec:rsqu}. 
Multiple runs of the algorithm failed to converge to a single geodesic: multiple local optima existed with similar values of the objective function. 
This suggests, by analogy with the Euclidean case, that the data are `isotropic' in treespace with no distribution around a particular geodesic direction.

\section{Discussion}

\subsection{Computational issues}

Construction of principal geodesics in treespace is computationally demanding, in most part due to the huge number of possible topologies for the end-points of geodesic segments. 
Since our algorithm is stochastic it is not possible to give an overall algorithmic complexity, but each evaluation of the objective function~$\eqref{equ:obj}$ has order $O(N^4\times n)$ where $N$ is the number of taxa and $n$ the number of data points. 
Calculation of the objective can be distributed across parallel processors simply by splitting up the data set into subsets, and so the algorithm speed increases linearly with the number of processors. 
The run times on a standard desktop computer were approximately $25$ minutes for the chaperonin data set and $2$ hours for the bootstrap sample of trees to obtain convergence from a single starting point. 
These times were obtained using $4$ cores on a Intel Core i$7$ CPU running at $2.93$GHz with $8$Gb memory.
However, very strict convergence criteria were used for the examples when preparing this article. 
In general the value of the objective function decreases approximately exponentially as the algorithm proceeds, and the majority of the runtime was spent performing minor adjustments to the segment $\gamma$ close to the final value. 
Faster times could be obtained if the user accepted a greater degree of approximation. 
In practice, the algorithm is probably limited to work with samples containing fewer than $100$ taxa and $500$--$1000$ trees on a standard desktop computer. 
However, parallel computing resources would enable larger data sets, particularly in terms of the number of trees, to be analysed. 
The algorithm is implemented in java, and software is freely available from the web site given in the abstract. 
An important part of the software is a tool for visualizing geodesics in treespace. 
In a similar way to phylogenetic network diagrams, it enables users to relate different topologies associated with a sample. 
The user drags a slider to traverse the geodesic and view the corresponding smoothly changing phylogenetic tree. 
The collection of topologies along a given geodesic segment is not always compatible with a single rotational ordering of taxa in the plane.  
Finding a rotational ordering for each topology in such a way as to minimize changes in ordering as the geodesic is traversed is problematic; the geodesic viewer uses simple heuristics to obtain a reasonable solution. 

\subsection{Further research}\label{sec:future}

Summarizing samples and distributions on treespace is a fundamentally difficult problem due to the high dimensionality and non-Euclidean nature of treespace. 
Dimensional reduction, by fitting appropriate low-dimensional objects to samples of trees, is an obvious approach to adopt. 
This article has focussed on the construction of an analog of the first principal component of a sample, without consideration of the higher order components. 
Construction of the second order component, for example, requires an analogue of a plane or some other $2$-dimensional surface in treespace, but the theory of higher-dimensional surfaces in treespace has not yet been developed. 
One possibility, a natural extension of the geodesic segments considered in this paper, is to seek a configuration of three points in treespace and consider the projection of the data onto the convex hull of these points. 
However, it is possible that completely different objects -- not necessarily based on the geodesic geometry -- might form better descriptors of distributions of phylogenies. 
Principal components analysis has been described in this article in terms of a least squares procedure rather than by reference to a probabilistic model. 
In this way, the sum of squared distances $d_\perp^2=\sum d(x_i,P_\gamma(x_i))^2$ has essentially played the role of the likelihood of the data. 
The likelihood for multivariate normal models on $\R^k$ has exactly this form, and this leads to the description of standard PCA as a least squares procedure, as given in the introduction. 
We adopted the least squares approach in treespace by analogy, but a probabilistic model and a more fully developed theory of distributions on treespace would be highly desirable.



\bibliographystyle{IEEEtran}


\end{document}